# Grain-boundary structure and segregation in Nb$_3$Sn coatings on Nb for high-performance superconducting radiofrequency cavity applications

Jaeyel Lee[1,2], Zugang Mao[1], Kai He[1,3,6], Zu Hawn Sung[2], Tiziana Spina[2], Sung-Il Baik[1,4], Daniel L Hall[5], Matthias Liepe[5], David N Seidman[1,4], Sam Posen[2]

[1]Department of Materials Science and Engineering, Northwestern University, Evanston, IL, 60208, USA

[2]Fermi National Accelerator Laboratory, Batavia, IL, 60510, USA

[3]Northwestern University Atomic and Nanoscale Characterization Experimental Center, Evanston, IL, 60208, USA

[4]Northwestern University Center for Atom-Probe Tomography (NUCAPT), Evanston, IL, 60208, USA

[5]Cornell Laboratory for Accelerator-Based Sciences and Education, Cornell University, Ithaca, NY, 14853, USA

[6]Department of Materials Science and Engineering, Clemson University, Clemson, SC, 29634, USA

**Abstract**

We report on atomic-scale analyses of grain boundary (GB) structures and segregation in Nb$_3$Sn coatings on Nb, prepared by the vapor-diffusion process, for superconducting radiofrequency (SRF) cavity applications, utilizing atom-probe tomography, high-resolution scanning transmission electron-microscopy and first-principles calculations. We demonstrate that the chemical composition of Nb$_3$Sn GBs is correlated strongly with the diffusion of Sn and Nb at GBs during the coating process. In a sample coated with a relatively large Sn flux, we observe an interfacial width of Sn segregation at a GB of ~3 nm, with a maximum concentration of ~35 at.%. After post-annealing at 1100 ºC for 3 h, the Sn segregated at GBs disappears and Nb segregation is observed subsequently at GBs, indicating that Nb diffused into the Nb$_3$Sn GBs from the Nb substrate. It is also demonstrated that the amount of Sn segregation in a Nb$_3$Sn coating can be controlled by: (i) Sn flux; and (ii) the temperatures of the Nb substrates and Sn source, which may affect the overall kinetics including GB diffusion of Sn and Nb. An investigation of the correlation between the chemical compositions of GBs and Nb$_3$Sn SRF cavity performance reveals that the Nb$_3$Sn SRF cavities with the best performance (high-quality factors at high accelerating electric-field gradients) do not exhibit Sn segregation at GBs. Our results suggest that the chemical compositions of GBs in Nb$_3$Sn coatings for SRF cavities can be controlled by GB engineering and can be utilized to optimize fabrication of high-quality Nb$_3$Sn coatings for SRF cavities.

| **Nomenclature** | | $\mu_i$ | chemical potential of an element, i |
|---|---|---|---|
| | | $\nu_{GB}$ | GB internal energy |
| A | area of GB plane | $\rho$ | atomic density of Sn in Nb$_3$Sn crystal |
| **c** | rotation axis of disorientation | APT | Atom-probe tomography |
| $d$ | average thickness of a Nb$_3$Sn film of each coating | CSL (Σ) | Coincident-site lattice |
| $d_g$ | diameter of a grain | EBSD | electron backscatter diffraction |
| $E_{acc}$ | accelerating electric field | EDS | energy disperse spectroscopy |
| $E_{antisite}$ | anti-site substitutional energy | FIB | focused ion-beam |
| $E_{segregation}$ | segregation energy | Five DOFs | five macroscopic degrees of freedom of a GB |
| $E_{Nb_3Sn}$ | total internal energy of bulk Nb$_3$Sn | GB | grain boundary |
| $E_{Sn\ in\ GB}$ | Sn occupation energy in the GB | GGA | generalized gradient approximation |
| $E_{GB}^{TOT}$ | total internal energy of GB structure | HAADF | high-angle annular dark-field |
| $E_{Sn\rightarrow Nb}^{TOT}$ | total internal energy of GB structure with a Sn anti-site | HA-GB | high-angle GB |
| $E_{Sn\ in\ GB}^{TOT}$ | total internal energy of GB structure with an additional Sn in the core of GB | HR-STEM | high-resolution scanning transmission electron microscopy |
| G | geometric factor | LA-GB | low-angle GB |
| **n** | normal vector perpendicular to GB {hkl} plane | PAW | projector augmented wave potentials |
| n | the number of Sn atoms located at the GB | PBE | Perdew-Burke-Ernzerhof |
| R$_s$ | surface resistance | Q$_0$-factor | Quality-factor |
| $t$ | time of Nb$_3$Sn coating process | SRF | superconducting radio frequency |
| T$_c$ | critical temperature | SU | Structural Unit |
| Γ$_i$ | Gibbsian interfacial excess of an element, i | VASP | *Vienna ab initio simulation package* |
| θ | rotation angle around **c**-axis | | |

1. **Introduction**

Significant progress has recently been made towards the application of $Nb_3Sn$ films for superconducting radio-frequency (SRF) cavities for particle-accelerator applications [1, 2]. $Nb_3Sn$ has an A15 structure and high-quality thin-films (~2-3 μm) of $Nb_3Sn$ are successfully created on the surfaces of niobium (Nb) cavities by employing the Sn vapor-diffusion process. With the substantially higher critical temperature ($T_c$ ~18 K) of $Nb_3Sn$ compared to Nb [3-5], 1.3 GHz SRF cavities demonstrate a $Q_0$ (quality)-factor of ~$10^{10}$ at 4.2 K, where the $Q_0$-factor is defined by a geometric factor (G) divided by the surface resistance ($R_s$) of a cavity ($G/R_s$), at useful accelerating electric fields; a value much higher, ~10 times, than has been achieved previously with Nb SRF cavities for the same experimental conditions (frequency, temperature, etc.) [6-8]. The performance of these $Nb_3Sn$ cavities has, however, been limited to an accelerating field of 14-17 MV/m, which corresponds to a peak surface magnetic field of ~70 mT. In some cases, SRF cavities display a significant $Q_0$-slope (a degradation of the quality factor with increasing electric-field) starting at ~8 MV/m. The corresponding surface magnetic field is significantly smaller than the theoretical superheating field of $Nb_3Sn$, $H_{sh}$ ~400 mT, predicted on the basis of the RF superconductivity of an ideal surface [1, 9, 10]. This RF degradation is thought to be caused by crystalline imperfections in the surface layer of the $Nb_3Sn$ coating on the Nb cavity's internal surface [2].

The coherence length ($\xi$) of $Nb_3Sn$ is significantly smaller than that of metallic Nb ($\xi(Nb_3Sn)$ ~3 nm versus $\xi(Nb)$ ~50 nm) [1, 8, 11], potentially making $Nb_3Sn$ more vulnerable to small-scale crystalline imperfections in the $Nb_3Sn$ layer. Several studies have been performed to investigate the structural imperfections in $Nb_3Sn$ coatings, such, as abnormally thin-grains [12-14], compositional variations [15-17], and Sn-deficient $Nb_3Sn$ phases [1, 8, 18], which may be responsible for the $Q_0$-slope and quenching, indicating a loss of superconductivity. Crystallographic orientation relationships between the Nb substrates and the $Nb_3Sn$ films were measured and the $Nb_3Sn$/Nb heterophase interfaces are demonstrated to play a crucial role in the formation of thin-grains and Sn-deficient regions [12]. Another candidate for causing degradation of the superconducting properties of $Nb_3Sn$ SRF cavities is grain-boundaries (GBs). In superconducting magnets, imperfections are purposely engineered into $Nb_3Sn$ wires at GBs. Impurities, such as normal conducting copper [19] or insulating oxides [20, 21] may be added to degrade locally the superconducting properties (e.g., superconducting order-parameter and superconducting energy gap [11, 22]) associated with GBs, which may create an energetically favorable volume for the nucleation of vortices, where they become pinned. In contrast, the magnetic-field direction in SRF cavities changes on a nanosecond time-scale, leading to a significant dissipation of heat and a corresponding decrease in the quality factor if the flux penetrates into the superconductor (including penetration into the GBs) [23]. The potential importance of GBs in $Nb_3Sn$ SRF cavities has been extensively discussed [3, 8, 10, 24, 25], but the structure and chemical compositions of GBs in $Nb_3Sn$, prepared by the vapor-diffusion process for SRF cavity applications are unknown, and, therefore, the exact roles played by GBs remains unclear.

The control of GB segregation has drawn scientific interest [26-29]. For practical applications, it affects the physical properties of alloys and ultimately provides a way to design materials with desired physical and mechanical properties. Especially for ordered alloys, such as $Ni_3Al$ [27, 30], $Ni_3Si$ [31], and CoAl [32], the overall composition is critical for segregation behavior at GBs, as stoichiometry away from the ideal concentrations of the alloys generates anti-site point defects, which tend to segregate at GBs to reduce the GB's total free-energy [27, 31]. This may provide opportunities for controlling the chemical compositions of GBs in ordered-alloy systems, $Nb_3Sn$, under the non-equilibrium vapor-diffusion condition. GB engineering of structures and chemical compositions of $Nb_3Sn$ GBs can open the possibility of improving the superconducting properties of $Nb_3Sn$ SRF cavities [33-36]. Clarification of this problem provides the motivation for our current studies.

Herein, we report the results of atomic-scale analyses of the chemical compositions and crystallographic structures of GBs, utilizing high-resolution scanning transmission electron microscopy (HR-STEM), atom-probe tomography (APT), and first-principles calculations at 0 K. We find that the chemical composition of $Nb_3Sn$ GBs is highly correlated with diffusion of Sn and Nb at GBs and it can be controlled by optimizing growth procedures. The results reveal Sn segregation at GBs of some $Nb_3Sn$ coatings, with a measured Gibbsian interfacial excess of 10-20 Sn atom/$nm^2$ caused by a continuous supply of Sn atoms from an external source, which diffuse along GBs. After post-annealing at 1100 °C for 3 h, the Sn segregation disappears and Nb segregation is then observed at GBs because the segregated Sn atoms diffuse away from the GBs and concomitantly Nb diffuses into the GBs from the Nb substrate. The amount of Sn segregation is controlled by two factors: (i) flux of Sn atoms; and (ii) the temperatures of the Nb substrate and Sn sources. The level of Sn segregation is the result of a series of kinetic processes. A correlation between the chemical compositions of $Nb_3Sn$ GBs and the performance of $Nb_3Sn$ SRF cavities are investigated. Our results demonstrate that GB engineering of GBs in $Nb_3Sn$ can be utilized to fabricate high-quality $Nb_3Sn$ coatings on Nb for SRF cavities.

## 2. Methodologies

2.1. Experimental procedure

The Sn vapor-diffusion process was utilized to coat the Nb substrate of either disk-shaped coupons (10 mm diameter with a 3 mm thickness) or the inner surface of 1.3 GHz cavities with $Nb_3Sn$ films [1, 3, 8]. Niobium substrates, a Sn source, and a $SnCl_2$ nucleating agent were placed in a vacuum furnace ($< 10^{-6}$ Torr). Two different growth conditions were employed for the $Nb_3Sn$ coatings to control GB segregation in $Nb_3Sn$ films; processes A and B, **Fig. 1**. For process A, the furnace temperature was initially increased to 500 °C to create nucleation sites on a Nb surface and the Sn source was concurrently heated to 500 °C. Then the temperature of the Nb substrate was increased to 1100 °C to initiate formation of the $Nb_3Sn$ phase on a Nb surface. During the coating process, the Sn source was maintained at 1200 °C, so that sufficient Sn was provided continuously to the top surface of $Nb_3Sn$. The Sn then diffused into the Nb sample. The temperature profile of the Sn source in **Fig. 1** was determined using a thermocouple close to the Sn source. The calibration procedure to estimate the actual temperature of the Sn source indicated that the real temperatures of the Sn source for processes A and B are ~1200 °C and 1220 °C, respectively [37]. The source contained 3.0 g Sn, which is more than the amount required for the desired film thickness of ~2 μm, and hence a Sn vapor is supplied continuously during the coating process. In process B, the Nb$_3$Sn coating was performed with both the Nb substrate and Sn source at an ~20 °C higher temperature, **Fig. 1**, which was achieved by placing the Nb substrates and Sn source close to the heating source of the furnace. Other differences are: (1) the source consisted of only 2.5 g Sn, so that together with the ~37% increase in the Sn evaporation rate, caused by elevating the temperature of the Sn source from 1200 to 1220 °C [38]; (2) all of the Sn source evaporated by the middle of the coating process and therefore, Nb$_3$Sn coatings experienced some annealing at the end of the coating procedure [8]. Previous data on the evaporation rate of the Sn source suggest that under these conditions, Nb$_3$Sn coatings may experience an annealing effect for 30 min to 60 min without an external supply of Sn.

The microstructural features of the Nb$_3$Sn samples were systematically evaluated employing the correlative use of a scanning electron-microscope (SEM), TEM, and APT. A thin TEM foil and an APT nanotip were prepared using a 600i Nanolab Helios focused ion-beam (FIB) microscope; the samples were sputter thinned with a 30 kV Ga$^+$ ion-beam at 87 pA, and then slowly sputter thinned utilizing 5 kV Ga$^+$ ions at 28 pA, to remove the radiation-damaged layers caused by the 30 kV Ga$^+$ ion-beam. HR-STEM imaging in conjunction with energy dispersive X-ray spectroscopy (EDS) were performed using an aberration-corrected JEOL ARM 200. For APT analyses, a Cameca LEAP4000X Si and a LEAP5000XS were utilized. Nanotips were irradiated with 30 pJ ultraviolet laser (wavelength = 355 nm) pulses at a pulse repetition rate of 250 to 500 kHz, with a detection rate of 0.5 to 1.0%. 3-D reconstructions and analyses were performed using Cameca's data analysis program, IVAS 3.8.1 (Cameca, Madison, WI).

2.2. Computational details

The first-principles calculations utilized in this research employ the plane-wave pseudopotential total-energy method as implemented in the *Vienna ab initio simulation package* (VASP). We used projector augmented wave (PAW) potentials [39, 40] and the generalized gradient approximation (GGA) of Perdew-Burke-Ernzerhof (PBE) for exchange-correlation [41]. Unless otherwise specified, all structures were fully relaxed with respect to volume, as well as to all cell-internal atomic coordinates. We considered carefully the convergence of results with respect to energy cutoff and k-points. A plane-wave basis set was employed with an energy cutoff of 600 eV to represent the Kohn-Sham wave functions. The summation over the Brillouin zone for the bulk structures was performed on a 12×12×12 Monkhorst-pack k-point mesh for all calculations. The magnetic spin-polarized method was applied in all the calculations. The lattice parameter of Nb$_3$Sn was calculated to be 5.332 Å, which is in excellent agreement with an experimental value of 5.289 Å [42]. The [1$\bar{2}$0]-tilt Nb$_3$Sn GB structures were constructed using the interface builder subroutine in the MedeA software package. The GB supercell had a total of 192 atoms, which contained two GBs.

### 3. Experimental results

3.1. TEM/APT analyses of Nb$_3$Sn grain-boundaries (GBs)

**Figs. 1(a,b)** display SEM and HAADF-STEM micrographs of typical ~2 μm thick Nb$_3$Sn coatings on Nb, prepared by the vapor-diffusion process A. The Gibbsian interfacial excesses of solute atoms and disorientations of thirty GBs detected in six Nb$_3$Sn samples were characterized systematically: using HR-STEM, EDS, APT, and transmission electron-backscatter diffraction (EBSD). The five macroscopic degrees of freedom (DOFs) of a GB are: (i) rotation axis, **c** (two DOFs), which is a unit vector; (ii) rotation angle, **θ**, about **c** (one DOF); and (iii) the GB plane as given by **n,** which is a unit normal vector to this (hkl) plane (two DOFs). These five DOFs were determined for four selected [1$\bar{2}$0] tilt-GBs among the thirty GBs observed to understand the effects of the five-macroscopic DOFs on the level of Sn segregation as determined by the Gibbsian excess, $\Gamma_i$. The different levels of Sn segregation, 10 to 20 atom/nm$^2$, are summarized in **Table 1**.

The chemical compositions of the GBs were determined by APT and the results are displayed in **Fig. 2**. The mass-spectrum of an APT analysis for a 3-D reconstruction and the identification of each peak are displayed in **Fig. S1**. Three-dimensional (3-D) reconstructions of the elements of one of the Nb$_3$Sn samples (A5), prepared by process A, is displayed in **Fig. 2(a)**, which presents a GB along the vertical direction. Iso-concentration surfaces for Nb = 70 at.%, Sn = 30 at.%, and N = 0.3 at.% are included in the 3-D reconstructions of each element [43], which indicate GB segregation of Sn and nitrogen (N). A 1-D-concentration profile across the GB plane is presented in **Fig. 2(b),** which indicates a maximum Sn concentration at a GB of 35 at.% and a Nb concentration of 65 at.%, which yield a value of $\Gamma_{Sn} = 18.1 \pm 1.7$ atom/nm$^2$ for the Gibbsian interfacial excess of Sn [44, 45]. The N concentration profile in **Fig. 2(c)** indicates that N also segregates at the same GB with an interfacial excess of $\Gamma_N = 1.6 \pm 0.1$ atom/nm$^2$. The full-width of the segregated region is estimated to be $3.0 \pm 0.2$ nm, where the width of Sn segregated region is defined by the full-width of the Sn concentration profile assuming that the Sn concentration profile follows a Gaussian function. Additional correlative TEM-APT analyses were performed for a GB from another sample (A10) prepared by process A, **Fig. 3**, with the GB plane located along the horizontal direction. In this case, the Gibbsian interfacial excess of Sn ($\Gamma_{Sn}$) has the value $11.4 \pm 0.8$ and $\Gamma_N$ is $0.5 \pm 0.1$ atom/nm$^2$ at the GB; these values are slightly smaller than those of the GB in sample A5. The full-width of the segregated region of sample A10 is $2.7 \pm 0.2$ nm.

HR-STEM imaging of selected [1$\bar{2}$0]-tilt GBs reveal the detailed atomic-structure of a GB in Nb$_3$Sn. **Fig. 4** presents a representative high-angle [1$\bar{2}$0]-tilt GB in Nb$_3$Sn. Owing to the orientation relationships (ORs) of the Nb$_3$Sn/Nb heterophase interfaces, some Nb$_3$Sn grains form [1$\bar{2}$0] tilt-GBs with different tilt-angles [12]. The structures and chemical compositions of GBs in Nb$_3$Sn coatings were also analyzed by atomically-resolved HR-STEM imaging and EDS mapping. HR-STEM micrographs of the high-angle (HA) tilt-GBs were recorded along the [1$\bar{2}$0]-zone-axis, revealing their atomic-structures. For the high-angle GB (HA-GB) of **Fig. 4(a)**, the Nb$_3$Sn grains are disoriented by $\theta = 46.6°$ about the [1$\bar{2}$0]-zone-axis (**c**), while the normal vector (**n**) to the GB plane was determined to be [215]. The coincident-site lattice (CSL) sites between two grains are indicated by yellow dotted circles, which appear every three (210) planes of the GB, implying a $\Sigma = 3$ coincidence site-lattice (CSL) [46]. The filtered HR-STEM micrograph of the HA-GB in **Fig. 4(b)** reveals structural-units (SUs), which are the cores of GB dislocations, as indicated by yellow pentagons [46, 47]. The HR-STEM image of the same GB with a GB plane normal, **n** = [21$\bar{1}$], which is rotated by 90° from the previous GB plane, (215), is presented in **Fig. S2**: it is also a $\Sigma = 3$ GB.

HR-STEM EDS measurements of Nb and Sn were also performed through the plane of the GB, revealing that the Sn concentration is 32 at.% Sn at the GB, **Fig. 4(b)**, compared to ~25 at.% Sn within the grains. The Gibbsian interfacial excess ($\Gamma_{Sn}$) was estimated to be 14.5 ± 2.3 atom/nm$^2$.

3.2  Effect of grain-boundary (GB) structure on segregation of Sn at a GB

Additional $[1\bar{2}0]$-tilt-GBs with different tilt angles (θ), were analyzed to investigate correlations with GB-structure and Sn-excesses at GBs. For a low-angle grain boundary (LA-GB), **Fig. 5(a),** the HAADF-HR-STEM micrograph displays two Nb$_3$Sn grains tilted by θ = 4.7° about the $[1\bar{2}0]$-axis (**c**), with the normal vector (**n**) to the GB plane being [001]. EDS measurements across the GB plane reveal a Gibbsian interfacial excess of Sn equal to $\Gamma_{Sn}$ =16.0 ± 6.4 atom/nm$^2$. Another $[1\bar{2}0]$-tilt-GB with θ = 17.6° is displayed in **Fig. 5(b)**, for which $\Gamma_{Sn}$ =18.5 ± 7.8 atom/nm$^2$ as determined by an EDS measurement. We note that the Gibbsian interfacial excess of Sn at the GBs is similar for both the HA-GB (14.5 ± 2.3 atom/nm$^2$) and LA-GB (16.0 ± 6.4 atom/nm$^2$), F**igs 4 and 5**, within experimental error. This implies that there is another factor that determines the level of GB-segregation of Sn in addition to the five macroscopic DOFs and their atomic structures. Tin segregation was analyzed systematically at additional GBs and the results are summarized in **Fig. 5(c)** and **Table 1**. The disorientation angles and axes of the GBs were determined by HR-STEM or transmission EBSD [48-50], **Fig. 5(c)**, and the chemistry was analyzed by HR-STEM EDS. When the values for the Gibbsian interfacial excesses of Sn at the GBs ($\Gamma_{Sn}$) are plotted as a function of the disorientation angle there isn't a clear correlation among the disorientation angles of the two grains at a GB and the level of Sn segregation.

3.3  Effects of the Sn-flux and temperatures of the Nb-substrates and Sn-source on Sn-segregation at GBs

**Fig. 6** displays the changes of the interfacial excesses of Sn as a function of the Sn flux and temperatures of the Nb substrates and Sn source. We note that the value of the average Sn flux is employed because there isn't an accurate method to measure the Sn flux during the coating process. The average Sn flux was estimated from the average Nb$_3$Sn film thicknesses and the coating times, and defined as $\rho d/t$, where $\rho$ is the atomic density of Sn in a $Nb_3Sn$ crystal, $d$ is the average thickness of a $Nb_3Sn$ film for each coating, and $t$ is the time for the $Nb_3Sn$ coating process. This plot yields a qualitative trend of Sn segregation at GBs, which depends on the Sn-flux. We investigated excess Sn at GBs prepared by two different processes: **Process A**: Sn and $SnCl_2$ are co-deposited on Nb substrates at T = 1100 ºC for 4 h with the Sn source at 1200 ºC for the first 3 h. The 3.0 g Sn source is not completely evaporated during process A, as there is some residual Sn deposition during the last hour with the Sn source at 1100 ºC; **Process B:** $Nb_3Sn$ samples are coated at a slightly elevated temperature, by 20 ºC for both Nb substrates (T = 1120 ºC), for 4 to 5 h and the Sn source (T = 1220 ºC) for the first 3 h. We think that the 2.5 g Sn source is evaporated completely by the middle of the coating process due to the ~37% increase of the Sn evaporation rate in process B compared to process A [8]; therefore, annealing occurs in the absence of a flux from the Sn source, for 30-60 min at the end of the coating process, **Fig.1(c)**.

Eight $Nb_3Sn$ samples were prepared by process A with different average Sn fluxes, which were analyzed by TEM. Tin excesses at the GBs were characterized utilizing EDS in the STEM mode for 1 to 4 GBs per sample. **Fig 6** presents the average values of Sn segregation at the GBs (Sn excess at GBs) of each sample, indicating that the Sn excesses are correlated with the average Sn flux. Tin segregation wasn't observed at the GBs in the low average Sn-flux samples, but there was evidence of an increase in the medium Sn-flux (100-150 Sn atom/$nm^2$ min) samples. The level of the Sn excess is saturated at ~18 atom/$nm^2$, with a corresponding maximum value of the GB concentration of ~35 at.% Sn.

Three samples coated by process B are displayed as red-dots in **Fig. 6** and the chemistry of 3 to 4 GBs were analyzed per sample utilizing STEM-EDS. Tin segregation wasn't observed at GBs coated according to process B. Tin depletion (or Nb segregation) was, however, observed for some GBs. The causes for the differences in the chemical compositions at GBs among samples produced by processes A and B are discussed in Section 4.2 and 4.3.

3.4 Effects of annealing on the Sn segregations at GBs

To identify the effect of annealing for 30-60 min in process B, annealing experiments were conducted for a Nb$_3$Sn sample (A5), which exhibited Sn segregation at GBs, **Fig. 2**. The Nb$_3$Sn sample with Sn segregation (A5) was annealed at 1100 °C for 3 h in the same vacuum furnace (< 10$^{-6}$ Torr) and two GBs of the annealed Nb$_3$Sn sample were analyzed using HR-STEM EDS. Indeed, Sn segregation disappears and Nb segregates at GBs, probably due to the diffusion of Nb from the Nb substrates, **Figs. 7** and **S3**, along GBs. The level of Nb segregation is 20 to 22 ±3 Nb atom/nm$^2$ for each GB and the full-width is ~5 ±1 nm for both GBs, which is a higher level and a wider full-width than for Sn segregation at GBs. The difference of Sn concentrations between the grain-interiors and the GB is ~9 at.%, which implies that the concentration of Sn at GBs decreases to ~16 at.%.

3.5 Segregation energy for Sn at a GB in Nb$_3$Sn

The GB structural model of a symmetrical Nb$_3$Sn 46.6°/[1$\bar{2}$0] tilt-GB with the GB-plane (21$\bar{1}$), **Fig. S.1**, was constructed for the purpose of first-principles calculations and is displayed in **Fig. 8(b)**. The GB was fully relaxed and the GB internal energy, $v_{GB}$, is determined using:

$$v_{GB} = (E_{GB}^{TOT} - 2E_{Nb_3Sn})/2A \tag{1}$$

Where $E_{GB}^{TOT}$ is the total internal energy of the GB structure displayed in **Fig. 8**, $E_{Nb_3Sn}$ is the total internal energy of one side of the bulk Nb$_3$Sn, and A is the area of the GB. The calculated GB internal energy at 0 K is 536 mJ/m$^2$.

For the case with an excessive amount of Sn (>25 at. %), the calculations were performed for three different locations of an extra Sn atom; two locations at the core GB sites with large free-volumes (labeled 1 and 2 in **Fig. 8(b)**), where the free-volume is defined as an excess volume attributed to any crystal defect [51], and one Nb atomic site in the bulk material to form an anti-site (labeled as number 3 in **Fig. 8(b)**) point-defect.

First, the anti-site substitutional energy of Sn for site 3 is calculated employing:

$$E_{antisite} = (E^{TOT}_{Sn \to Nb} + \mu_{Nb}) - (E^{TOT}_{GB} + \mu_{Sn}) \qquad (2)$$

Where $E^{TOT}_{Sn \to Nb}$ is the total GB energy with a Sn anti-site point-defect, and $\mu_{Nb}$ and $\mu_{Sn}$ are the chemical potentials for pure Nb and Sn atoms. The calculated Sn anti-site energy is 0.635 eV/atom.

Next, the Sn occupation energy at core GB sites (sites 1 and 2) is calculated using:

$$E_{Sn\,in\,GB} = [E^{TOT}_{Sn\,in\,GB} - (E^{TOT}_{GB} + \mu_{Sn})]/n \qquad (3)$$

Where $E^{TOT}_{Sn\,in\,GB}$ is the total GB energy with a Sn atom situated in the core of a GB, site 1 or 2, as displayed in **Fig. 8**, and *n* is the number of Sn atoms located at a GB. The calculated Sn occupation energy at the core of a GB is -0.164 eV/atom at site 1 and -0.152 eV/atom at site 2. Tin occupation at a GB can stabilize the GB structure by decreasing the system's total energy at both the sites studied herein. Site 1 has a slightly more negative occupation energy than site 2 due to having a 24% larger free-volume, **Fig. 8**.

Finally, the driving force for excess Sn segregation from the bulk to a GB can be determined by calculating the segregation energy as follows:

$$E_{segregation} = E_{Sn\,in\,GB} - E_{antisite} \qquad (4)$$

The value of the segregation internal energy calculated at 0 K for our system is -0.799 eV/atom for site 1 and -0.787 eV/atom for site 2 at the GB, which are quite close to one another.

### 4. Discussion

4.1. Structures and chemical compositions of Nb$_3$Sn GBs

TEM and APT analyses of the atomic structure and chemical compositions of GBs in Nb$_3$Sn demonstrate that the full-width of the Sn segregated region at a GB is ~3.0 ±0.5 nm and the maximum Sn concentration is ~35 at.% (**Figs. 2-4**). As seen in **Fig. 4(b)**, periodic atomic arrays of structural units (SUs) (the cores of GB dislocations) are observed in the HR-STEM images. A possible sign of chemical ordering of segregated Sn at a GB is observed in a limited case, as indicated by the periodic bright-contrast effects of coincident sites at a GB of **Figs. 4(a)** and **S4**. This implies that there could be a higher concentration of Sn atoms in an atomic column in the core of a GB dislocation, agreeing with the results of our first-principles calculations. There is, however, no clear evidence for the presence of chemical ordering or non-equilibrium phases [52-55] at other GBs employing additional rigorous analyses using HR-STEM, and we conclude that most GBs form structurally sharp interfaces, **Fig. 5**.

Our first-principles calculations demonstrated that GBs are favorable sites for excess Sn in $Nb_3Sn$. As the formation energy of a Sn-anti-site defect is high, 0.635 eV [12, 56], Sn-anti-site defects are most likely to segregate at GBs to reduce the grain-boundary energy of $Nb_3Sn$. When a Sn-anti-site defect moves from the bulk (site 3) to a core GB site (1 and 2 in **Fig. 8(b)**), the energy is reduced by ~0.8 eV/atom, which is the segregation energy of Sn at the cores of GBs. Excess Sn atoms may first fill the core GB sites with large free volumes but since these sites are limited, it may then begin to occupy a Nb sublattice site near the GBs, within the ~3 nm width, after filling the GB core-sites. The segregation energy of a Sn atom at the Nb sublattice near GBs may be smaller than that at the core sites of a GB, because segregated Sn forms a Sn anti-site defect near a GB. This reduction in segregation energy means that after filling the core GB sites, Sn segregation at GBs becomes more costly energy-wise, which may explain the observation that saturation of Sn segregation occurs at 15~20 atom/$nm^2$ even with a high Sn-flux, **Fig. 6**. An estimated 15 atom/$nm^2$ of Sn excess in ~3 nm surrounding a GB generates about one Sn anti-site defect in every two unit-cells of $Nb_3Sn$. This results in a substantial increase in internal energy, which is probably insufficient to overcome the nucleation barrier to transform to other phases.

4.2. Origin of Sn segregation at GBs in $Nb_3Sn$

In this section, we consider the causes of Sn segregations at GBs based on thermodynamic and kinetic reasons. There are two thermodynamic factors: (i) GB structure; and (ii) stoichiometry of $Nb_3Sn$ associated with the formation energies of Sn and Nb antisites. The GB structures are highly meta-stable, so following J. W. Cahn's *ansatz* that the five macroscopic DOFs of a GB are thermodynamic state variables, which determine the structure of a GB, we studied GB structure and its relationship to segregation [52]. Cahn assumed that the three microscopic DOFs are relaxed to their equilibrium values, which may or may not be the case for our situation. For the kinetic origins, Sn (and Nb) diffusion with or without a Sn-flux during the coating process is considered.

Firstly, it is noteworthy that GB segregation for the vapor-diffused $Nb_3Sn$ layer is not correlated with the structures of the GBs as represented by the five macroscopic DOFs. This is in contrast to the situation for an equilibrium state, where the interfacial excesses of solute are determined by the structures of the GBs, represented by the five macroscopic DOFs [33, 57]. LA-GBs contain smaller dislocation densities than HA-GBs, and therefore, in general, should exhibit smaller values of the Gibbsian excesses of solute atoms [34]. We have, however, observed no detectable differences in the amount of Sn segregation at LA-GBs and HA-GBs, **Figs. 4** and **5**.

Another thermodynamic origin that affects segregation behavior at GBs in $Nb_3Sn$ is the stoichiometry of $Nb_3Sn$ associated with the formation energies of Sn and Nb antisite point defects. $Nb_3Sn$ is an ordered alloy, where the segregation behavior of solutes is affected by the stoichiometry of the ordered intermetallic phase [27, 30-32, 58, 59]. For ordered alloys, the chemical composition of the GBs is related to the formation of anti-site defects in the interior of grains and their segregation at GBs to reduce GB energies. First-principles calculations at 0 K demonstrate that excess Sn inside $Nb_3Sn$ grains has a high driving-force for segregation at GBs, ~0.8 eV/atom. If the Sn concentration in $Nb_3Sn$ is <25 at.% Sn, Nb is the excess element and the excess Nb atoms then segregate preferentially at GBs to reduce their energy. We tried to analyze the effect of the bulk composition of $Nb_3Sn$ on Sn segregation at GBs employing APT/STEM-EDS analyses and strong correlations aren't observed, **Fig. S5**: some $Nb_3Sn$ grains that exhibit Sn-deficient regions display Sn-segregation at GBs and some possible Sn-rich $Nb_3Sn$ grains with ~25.5 at.% Sn do not display Sn segregation at GBs.

Lastly, the role of kinetics is discussed. In **Fig. 6**, the level of Sn segregation at GBs changes with different values of the average Sn flux. Higher Sn fluxes cause larger levels of Sn segregation at GBs, **Fig. 6**. This indicates that Sn diffusion along GBs during the coating process, with an external Sn-flux, plays an important role in Sn segregation at GBs. With an external Sn source, there is a continuous Sn flux to $Nb_3Sn$ coatings from the top surface and GBs in $Nb_3Sn$ are the main paths for Sn diffusion into $Nb_3Sn$ and Nb substrates. Alternatively, annealing a Nb$_3$Sn sample, without an external Sn source, induces diffusion of the segregated Sn atoms from the GBs and Nb diffusion into the GBs from the Nb substrates, which results in Nb segregation at GBs, **Fig. 7**. The outcome that Sn-segregation is controlled by annealing with or without a Sn-flux is strong evidence that the Sn segregation is caused by Sn-diffusion along GBs, due to the external Sn-flux, which is a kinetic effect.

Segregation at GBs in Nb$_3$Sn in the current study (especially for process A) was observed in a non-equilibrium state, with a continuous supply of Sn atoms from the vapor phase. Short-circuit diffusion of Sn and Nb along GBs and interfacial reactions at the Nb$_3$Sn/Nb interface continue until the termination of the coating period for both processes A and B. In this non-equilibrium state of the Sn-vapor diffusion coating process, the effects of kinetics on GB segregation in Nb$_3$Sn films may dominate over thermodynamic factors and it may rationalize the absence of clear effects of GB structure and the stoichiometry of Nb$_3$Sn on the level of Sn segregation, within the detection limit of STEM-EDS and APT analyses.

We note, however, that even though the kinetics is the primary origin that determines the chemical compositions of GBs in Nb$_3$Sn during the coating process, the higher level of Nb segregation and the wider width of the Nb concentration profile at GBs (**Figs 7** and **S3**), compared to Sn segregation, is attributed partially to the thermodynamic origin: a smaller formation energy of a Nb anti-site defect (0.268 eV) compared to a Sn anti-site defect (0.637 eV). Nb$_3$Sn has a wide range of composition, ranging from 17 to 26 at.% Sn, which allows for variations in the chemical compositions of GBs in Nb$_3$Sn, especially on the Nb-rich side of the Nb-Sn phase diagram [60].

It is noteworthy that the level of Sn segregation is essentially a constant along the depth of GBs from the top-surface of Nb$_3$Sn to the Nb$_3$Sn/Nb heterophase interface, **Fig. 9**. This result indicates that GB diffusion of Sn is so rapid that there isn't a concentration gradient along a GB [61]. The short-circuit diffusivity of Sn along GBs ($D_{gb}$) is faster by more than three orders of magnitude than bulk diffusion in Nb$_3$Sn ($D_b$) at 1100 °C; the estimated root-mean-square (RMS) diffusion distance of Sn in bulk Nb$_3$Sn and in GBs at 1100 °C in 1 h ($\sqrt{4Dt}$) are ~200 nm and ~120 μm, respectively [12, 56]. The RMS bulk diffusion distance of Sn at 1100 °C, ~200 nm, is larger than the GB diameter ($\sigma = 3$ nm), but it is much smaller than the grain diameter ($d_g = $ ~2 μm). This implies that GB diffusion is the primary path for Sn diffusion in Nb$_3$Sn coatings and $D_{gb}$ for Sn in GBs in our Nb$_3$Sn coatings is classified as type-B diffusion, according to Harrison's classification scheme, where σ<<$\sqrt{Dt}$<<$d_g$ [62, 63], which is called the *leaky-pipe* model.

4.3. Controlling the chemical compositions of Nb$_3$Sn GBs by growth processes

The key finding of the current research is that Sn segregation at a GB can be controlled during annealing with or without a Sn-flux. Specifically, it is highly correlated with, and can be controlled by, two factors: (i) the Sn flux; and (ii) the temperature of the Nb substrate and Sn source as observed in processes A and B, **Fig. 6**. Tin segregation at GBs disappears and Nb segregates at the GBs of some Nb$_3$Sn coated films utilizing process B. The causes of the different chemical compositions of GBs in the Nb$_3$Sn coatings prepared by process B compared to process A are: (i) there isn't an external Sn flux during the last 30 to 60 min at the end of the coating period in process B; (ii) as a result, the Nb$_3$Sn coatings are annealed at 1120 °C for an extra 30 to 60 min and the segregated Sn atoms have sufficient time to diffuse into the Nb substrate to form Nb$_3$Sn. Longer annealing times, without an external Sn source, cause Nb diffusion from the Nb substrate into Nb$_3$Sn GBs, which leads to Nb segregation at GBs, **Fig. 7**. Variations of GB segregation in the presence or absence of an external vapor source have also been reported for the Cu-Bi system [29], where segregation of Bi at GBs in Cu is induced by an external Bi vapor-source, where the level of Bi segregation at a GB achieves a stationary-state. Once the external Bi vapor-source is removed the segregated Bi atom concentration at the GBs decreases until an equilibrium Bi vapor-pressure is established. Another difference in process B compared to process A is that the elevated temperatures of the Nb substrate and Sn source probably increase the interfacial reaction rate at the Nb$_3$Sn/Nb interface. Assuming the activation energy barrier for Nb$_3$Sn growth is ~230 kJ/mole [64], then increasing the temperature from 1100 °C to 1120 °C accelerates the growth rate of Nb$_3$Sn by ~34%, assuming the Nb$_3$Sn film growth follows an Arrhenius expression [64-66]. This increase of the interfacial reaction rate at the Nb$_3$Sn/Nb interface may consume ~34% more Sn atoms at Nb$_3$Sn/Nb interfaces, which helps to reduce Sn segregation at GBs.

### 4.4. Grain-boundary engineering for high-performance Nb$_3$Sn SRF cavities

The width of Sn segregation at a GB (~3 nm) is comparable to the superconducting coherence length of Nb$_3$Sn [2, 10, 23]. This suggests that Sn segregation at GBs can indeed possibly provide a path for magnetic flux penetration from the surface. Our research demonstrates that the chemical compositions of Nb$_3$Sn GBs for the Sn vapor-diffusion process can be fine-tuned utilizing an optimized growth-process procedure, **Fig. 6**. Thus, GB engineering of Nb$_3$Sn coatings provides a route for improving the superconducting properties of Nb$_3$Sn SRF cavities [33], which is the goal of our research.

One Nb$_3$Sn sample was cut directly from a high-performance Nb$_3$Sn cavity, Cornell cavity 2 (**Fig. 10(a)**) [67], to investigate the correlation between Sn segregation at GBs and surface resistance. When GBs from a high-performance Nb$_3$Sn cavity were analyzed by HR-STEM EDS and APT, there were no signs of Sn segregation, although a small amount of Sn depletion (or Nb segregation) of ~4 atom/nm$^2$ was observed at one GB, **Fig. 10(b)**. Another Nb$_3$Sn coated cavity, Fermilab cavity 3, was prepared using process B in an attempt to avoid Sn segregation. A witness sample was coated together with Fermilab cavity 3 by process B to simulate Nb$_3$Sn films in a cavity. STEM-EDS analysis of the GBs of the witness Nb$_3$Sn sample revealed a lack of Sn segregation at GBs. A Sn concentration profile through a GB is displayed in **Fig. 10(c)**. Furthermore, Fermilab cavity 3 demonstrated a high level of cavity performance, with a minimal Q-slope until an accelerating gradient of 15 MV/m.

Alternatively, Fermilab cavities 1 and 2, which experience degradation of the Q-factor at ~8 MV/m, do exhibit Sn segregation at GBs in the witness samples of the cavities, which are coated using the same growth condition to Fermilab cavities 1 and 2 (numbers 1 and 2, **Fig. 6**). The Sn excesses at the GBs are equal to 13.9 ± 1.8 and 12.3 ± 6.6 atom/nm$^2$ for samples 1 and 2, respectively. Our present results suggest that it is reasonable to postulate a correlation between Sn segregation at GBs in Nb$_3$Sn and the appearance of Q-slope or quench in a Nb$_3$Sn cavity. This could indicate that GB segregation of Sn increases the surface resistance in Nb$_3$Sn cavities and it is possible that the smaller Sn segregation at GBs may mitigate the degradation of Nb$_3$Sn cavity performances (Q-slope and quench of Nb$_3$Sn cavities).

We also investigated the possible effects of Sn-deficient regions and surface chemistry and to date significant roles aren't observed [68]. It is extremely challenging to isolate the effect of GBs on the SRF cavity performance from other factors, such as Sn-deficient grains, surface roughness, and surface chemistry, and this will require further investigations. Theoretical evaluations on the correlation between Sn segregation at GBs and superconducting properties is ongoing using density functional theory (DFT) calculations of the change of the critical temperature of $Nb_3Sn$ for different Sn concentrations and a time-dependent Ginzburg-Landau model to simulate the penetration of magnetic fluxes into the Sn segregated GB, which will be reported in separate articles [69-71]. For $Nb_3Sn$ magnet applications, there has been intensive research on the role of GBs as flux-pinning centers, demonstrating that GBs in $Nb_3Sn$ magnets are pinning-centers of magnetic flux owing to the short coherence length of $Nb_3Sn$ (~3 nm), which enhances the current density [72-76]. Present investigations suggest that GBs in $Nb_3Sn$ are probably relevant to SRF cavity applications as well, and they provide a pathway to control the detailed chemical compositions of GBs in $Nb_3Sn$ used for SRF cavity applications.

## 5. Conclusions

This article describes the use of atom-probe tomography (APT), high-resolution scanning transmission electron microscopy (HR-STEM), and first-principles calculations at 0 K, to perform systematic analyses of the chemical compositions and structures of grain boundaries (GBs) in $Nb_3Sn$ for superconducting radiofrequency (SRF) applications. Our results indicate a series of relationships among the growth kinetics, chemical compositions of grain-boundaries (GBs) and $Nb_3Sn$ SRF cavity performance.

- Grain-boundary short-circuit diffusion of Sn and Nb play an important role on the chemical compositions of GBs in $Nb_3Sn$, which can be controlled by optimizing the growth processes.
- Tin segregation is observed at GBs in some $Nb_3Sn$ coatings and it is attributed to Sn short-circuit diffusion along GBs, due to the continuous Sn flux from an external Sn source. APT and HR-STEM-EDS analyses reveal that 10-20 atom/$nm^2$ of excess Sn segregates at the GBs, within a ~3 nm full-width and a maximum Sn concentration of 35 at.% Sn.

- Tin segregation at GBs disappears after post-annealing at 1100 °C for 3 h and concomitantly Nb segregates at the GBs due to diffusion of the segregated Sn from the GBs and simultaneously Nb diffusion along the GBs from the Nb substrates.

- The amount of Sn segregation at a GB during the coating process is correlated with two factors: (1) Sn flux; and (2) the temperatures of the Nb-substrate and Sn-source. These parameters affect short-circuit Sn diffusion along GBs and probably interfacial reactions at $Nb_3Sn$/Nb interfaces, which determines the chemical composition of a GB.

- First-principles calculations at 0 K demonstrate that excess Sn in $Nb_3Sn$ segregates preferentially at GBs due to the high formation-energy of Sn-anti-site defects in bulk $Nb_3Sn$ (0.635 eV).

- An investigation of the correlation between Sn segregation at GBs and cavity performance demonstrates that high-performance $Nb_3Sn$ SRF cavities are probably achieved when *GBs in $Nb_3Sn$ coatings do not exhibit significant Sn or Nb segregation*.

- We have shown experimentally that studies of GBs in $Nb_3Sn$ are relevant to the $Q_0$ (quality)-factor of $Nb_3Sn$ SRF cavities as a function of the accelerating electric field (MV/m), **Fig. 10**, owing to the short coherence length (~3 nm) of $Nb_3Sn$.

- Implementation of the optimum process may facilitate the fabrication of high-performance $Nb_3Sn$ cavities. *Our results help, therefore, to improve the fabrication of high-quality $Nb_3Sn$ coatings for SRF cavity applications*.

**Acknowledgments**

We are grateful to Drs. Amir R. Farkoosh, Xiaobing Hu, Dieter Isheim, Shipeng Shu, Xingchen Xu, Xuefeng Zhou and Mr. Qingqiang Ren for valuable discussions, and Mr. Brad Tennis for assistance in processing $Nb_3Sn$ coatings. We also thank Profs. Thomas Arias, David A. Muller, and James P. Sethna for their valuable suggestions. This research is supported by the United States Department of Energy, Offices of High Energy. Fermilab is operated by the Fermi Research Alliance LLC under Contract No. DE-AC02-07CH11359 with the United States Department of Energy. This work made use of the EPIC, Keck-II, and/or SPID facilities of Northwestern University's NU*ANCE* Center, which received support from the Soft and Hybrid Nanotechnology Experimental (SHyNE) Resource (NSF ECCS-1542205); the MRSEC program (NSF DMR-1121262) at the Materials Research Center; the International Institute for Nanotechnology (IIN); the Keck Foundation; and the State of Illinois, through the IIN. Cornell's $Nb_3Sn$ coating program is supported by United States Department of Energy grant DE-SC0008431. Atom-probe tomography was performed at the Northwestern University Center for Atom-Probe Tomography (NUCAPT). The LEAP tomograph at NUCAPT was purchased and upgraded with grants from the NSF-MRI (DMR-0420532) and ONR-DURIP (N00014-0400798, N00014-0610539, N00014-0910781, N00014-1712870) programs. NUCAPT received support from the MRSEC program (NSF DMR-1720139) at the Materials Research Center, the SHyNE Resource (NSF ECCS-1542205), and the Initiative for Sustainability and Energy (ISEN) at Northwestern University.

**Table 1**. Summary of Gibbsian interfacial excesses ($\Gamma_i$) and the five macroscopic degrees of freedom (DOF) of selected [$1\bar{2}0$]-tilt-GBs.

| Number | $\Gamma_{Sn}$ (atom/nm$^2$) | Five macroscopic degrees of freedom (DOFs) | | Methods |
|---|---|---|---|---|
| | | Disorientation (**c**, θ) | GB plane (**n**) | |
| GB1-1 | 14.5 ± 2.3 | [$1\bar{2}0$], 46.6° | (215) | HR-STEM / EDS |
| GB1-2 | | [$1\bar{2}0$], 46.6° | ($21\bar{1}$) | |
| GB2 | 16.0 ± 6.4 | [$1\bar{2}0$], 4.7° | (001) | |
| GB3 | 14.6 ± 1.5 | [$1\bar{2}0$], 30.0° | (212) | |
| GB4 | 18.5 ± 7.8 | [$1\bar{2}0$], 17.5° | (215) | |

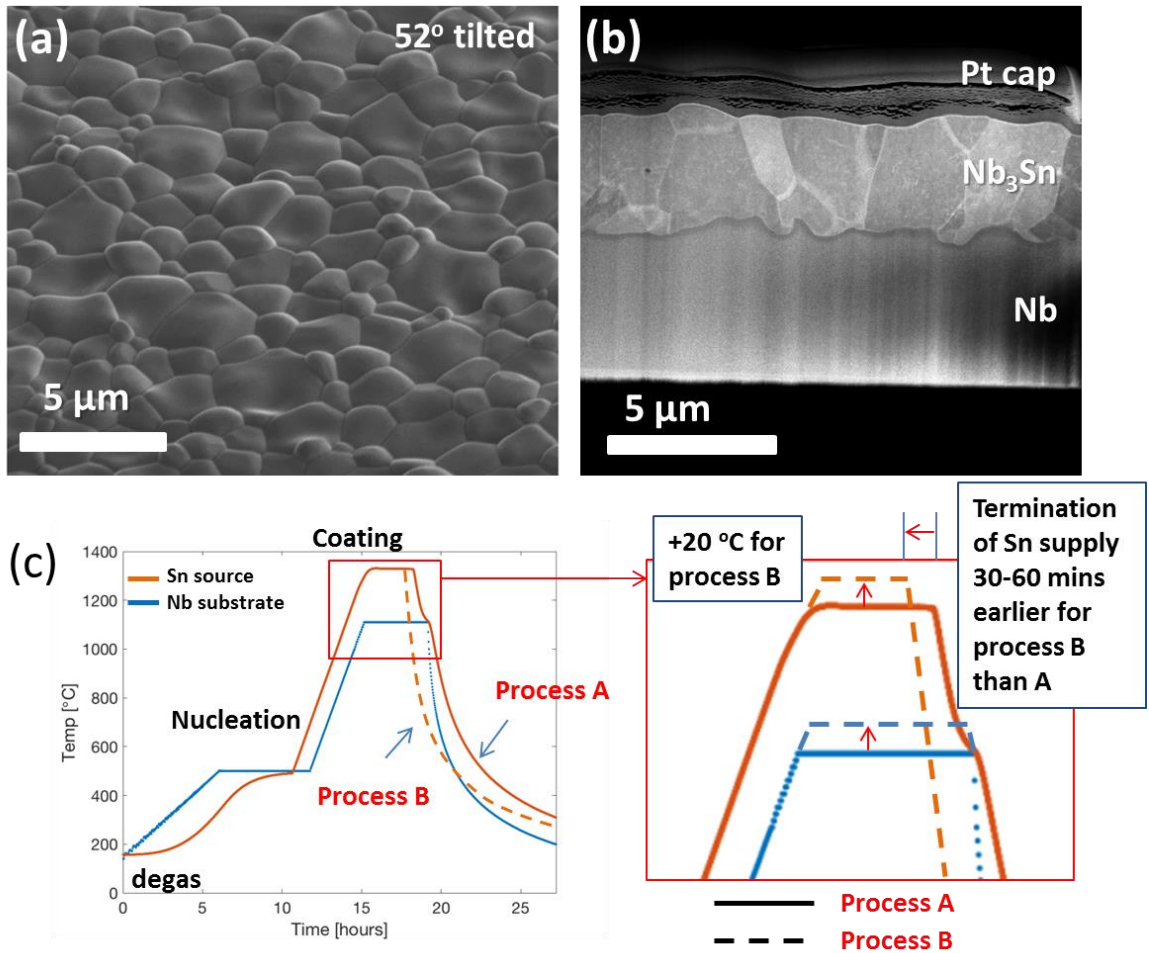

**Figure 1**. (a) SEM micrograph of the surface of a Nb$_3$Sn coating on Nb tilted from the optic axis of the electron beam by 52°; (b) HAADF-STEM cross-sectional micrograph of a Nb$_3$Sn coating on Nb; (c) Schematic temperature profiles of the Sn source and Nb substrates (furnace) for Nb$_3$Sn coatings on Nb employing processes A or B. For process B, the temperatures of the Sn source and Nb substrates are 20 °C higher than for process A. Additionally, for process B, the external Sn source is depleted at 30 to 60 min prior to the end of the coating period and therefore the Nb$_3$Sn coatings experience some annealing.

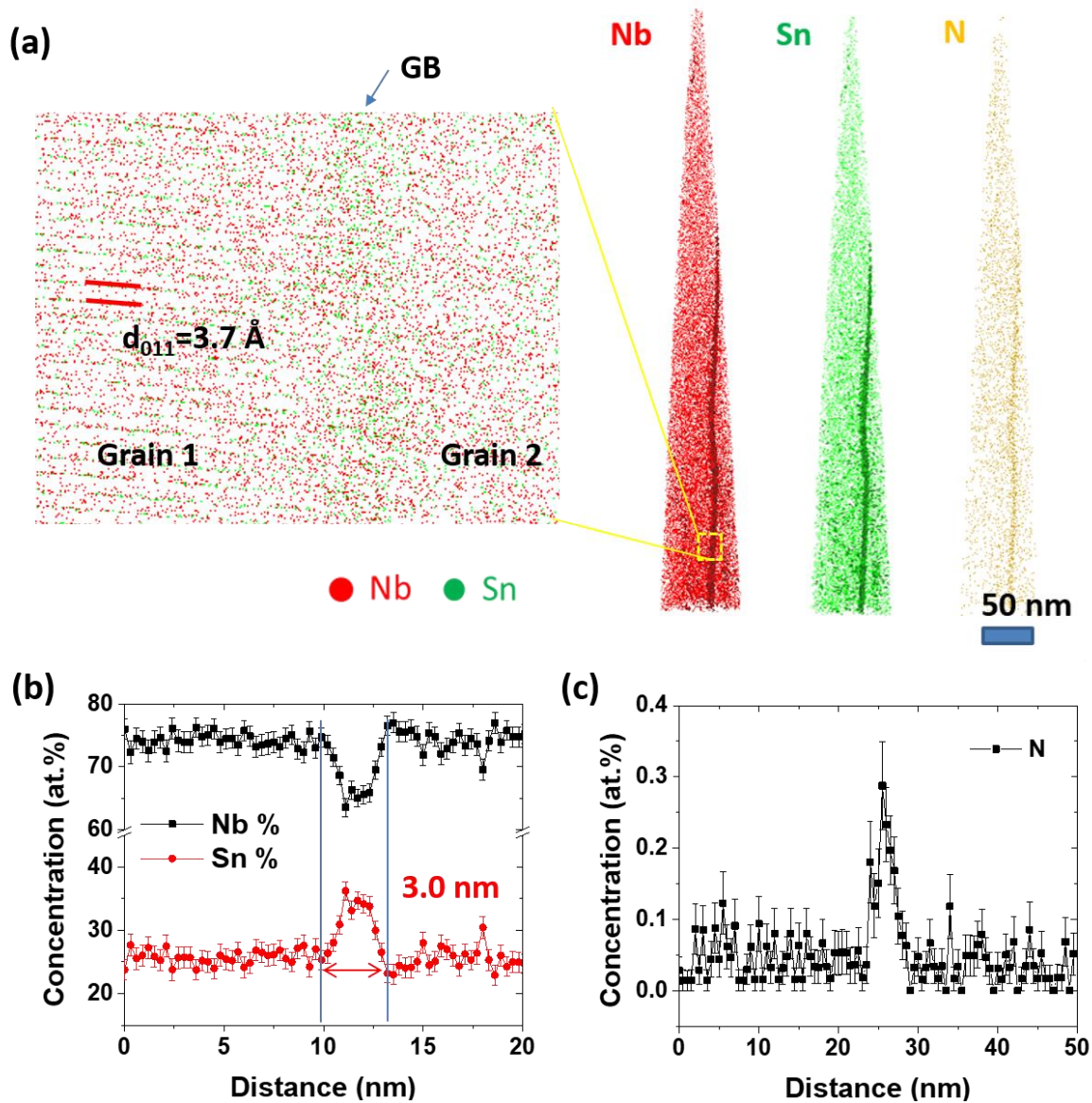

**Figure 2.** (a) 3-D reconstructed Nb, Sn, and N atomic distribution-maps. Iso-concentration surfaces of Nb (70 at.% Nb), Sn (30 at.% Sn), and N (0.3 at.% N) are indicated in the 3-D reconstructed maps of each element, upper right-hand figure. (b) Concentration profiles across a GB plane displaying Sn segregation and concomitantly Nb depletion. The measured full-width of the segregated zone is $3.0 \pm 0.2$ nm: the segregated zone is defined as the full-width of the Sn segregation and Nb depletion profiles as indicated by the two vertical blue-lines; (c) Nitrogen concentration profile across a GB.

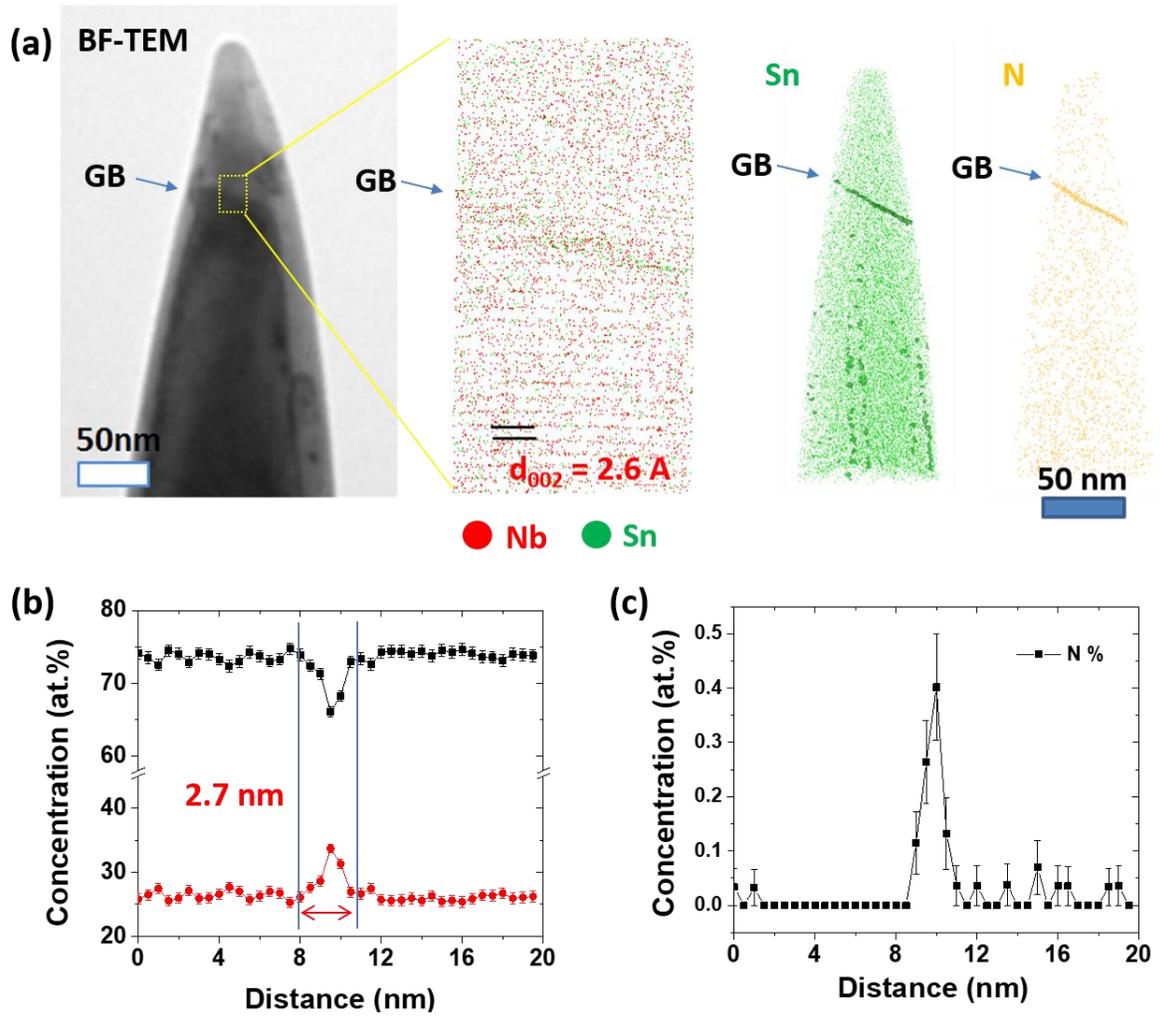

**Figure 3** (a) BF-TEM image of a nanotip of $Nb_3Sn$ for the APT experiment and corresponding 3-D reconstruction of Nb, Sn, and N atoms: traces of the GB plane are also indicated. (b) 1-D concentration profiles of Nb, Sn, and N across a GB displaying segregation of Sn and N at this GB. (c) Nitrogen (N) concentration profile.

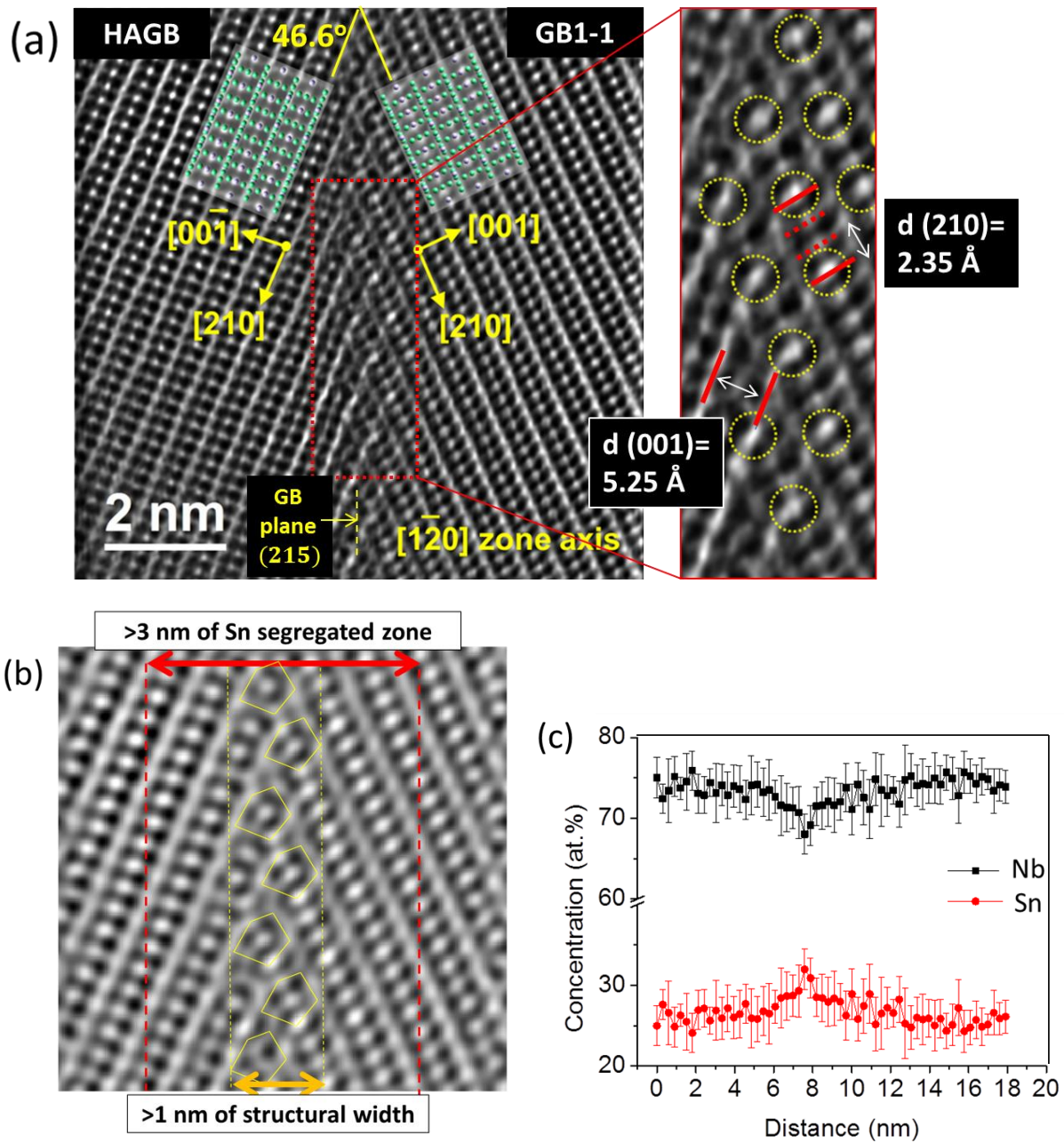

**Figure 4**. (a) Atomically resolved HR-STEM micrograph of a high-angle tilt-GB ($\theta = 46.6°$, **c**= [1$\bar{2}$0]) in Nb$_3$Sn, where the normal to the GB plane is **n** = [215]. CSLs at the GB are marked with yellow dotted-circles in the magnified image. (b) Filtered HR-STEM micrograph displaying a periodic array of structural units (cores of GB dislocations) of the HA-GB, indicated by yellow pentagons. (c) EDS concentration profiles

across the HA-GB demonstrating that the maximum Sn concentration at the GB is ~32 at.% Sn, which is higher than in the matrix, 25 at.% Sn. Note the depletion in the Nb concentration profile.

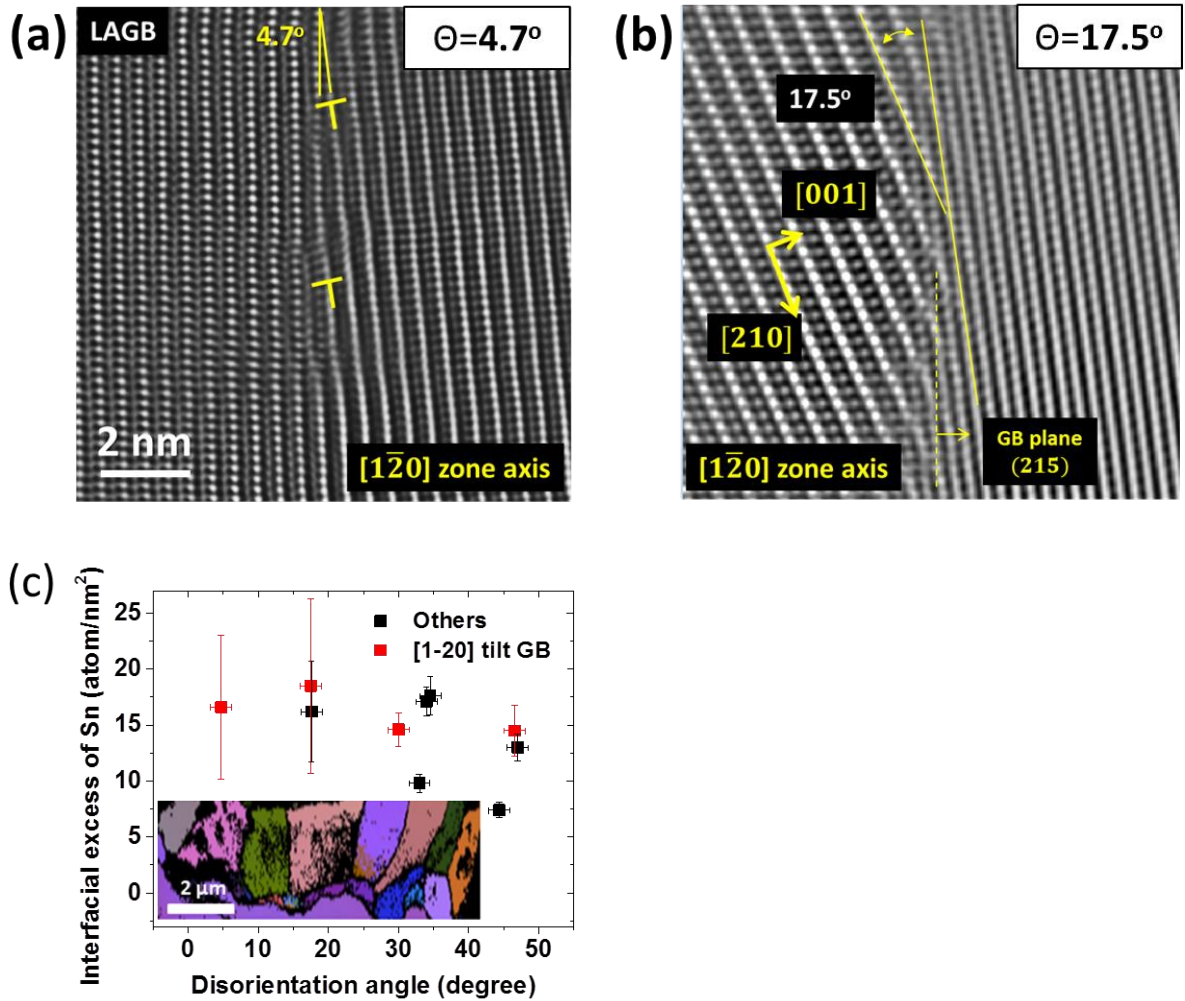

**Figure 5**. Atomically resolved HR-STEM images of $[1\bar{2}0]$-tilt-GBs in $Nb_3Sn$ with different tilt angles ($\theta$) and GB planes (**n**): (a) $\theta = 4.7°$/**n** = [001] and (b) $\theta = 17.5°$/**n** = [215]. (c) Plot of the Gibbsian interfacial excesses of Sn ($\Gamma_{Sn}$) as a function of the disorientation angle of GBs. The red-squares are for a $[1\bar{2}0]$ rotation axis with different tilt-angles and black-squares are for random rotation axes. A transmission-EBSD color map of a TEM sample is inset in this plot, which was used to identify the disorientation angles of the GBs.

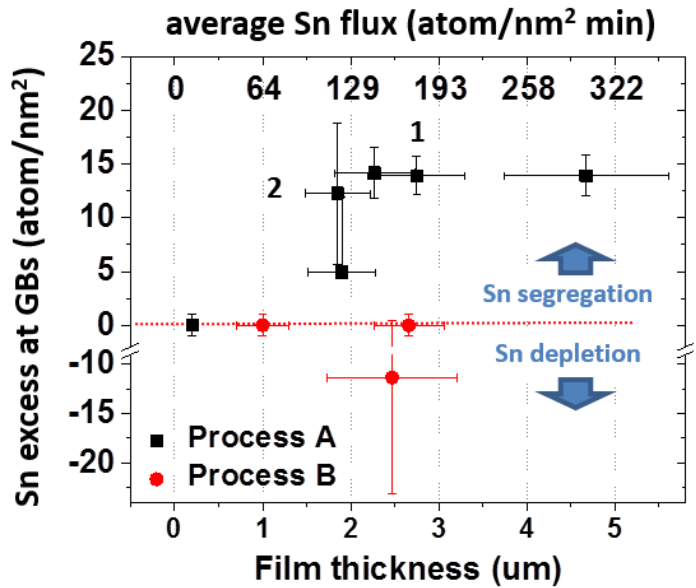

**Figure 6**. Gibbsian interfacial excesses of Sn at GBs in $Nb_3Sn$ films coated using processes A or B, which differ with respect to the average Sn flux and the temperatures of the Nb substrates. For process A, $Nb_3Sn$ films were deposited on Nb substrates at 1100 °C with the Sn source at 1200 °C; the Sn vapor was supplied during the entire coating process. For process B, the temperature of both the Nb substrate and Sn source were increased by 20 °C during creation of the $Nb_3Sn$ films; there wasn't, however, a Sn flux for the last 30-60 min of this process.

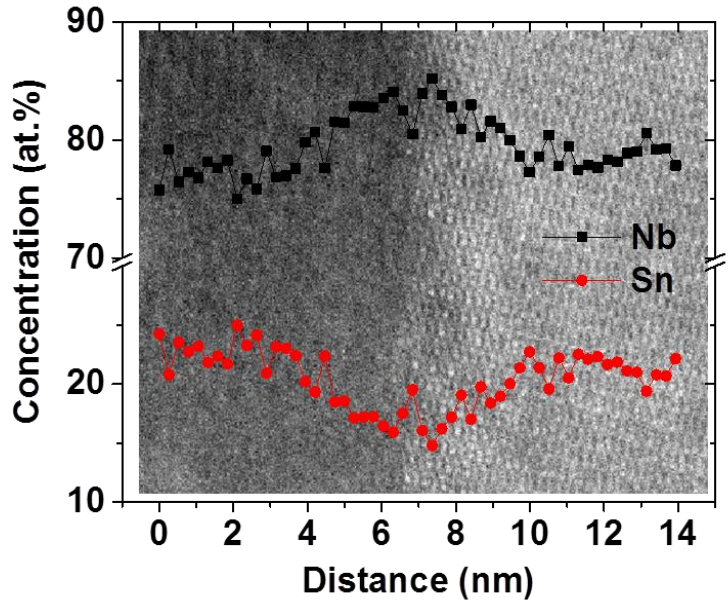

**Figure 7**. Concentration profiles across a GB and the corresponding HR-STEM micrograph of a Nb$_3$Sn sample, which originally exhibited Sn segregation at GB in the as-coated state (A5) and then was annealed at 1100 °C for 3 h. Tin segregation disappears and concomitantly Nb segregation is observed. The Gibbsian interfacial excess of Nb is estimated to be 22 ±3 atom/nm$^2$ and the full-width of the Nb segregated region is ~5 ±1 nm.

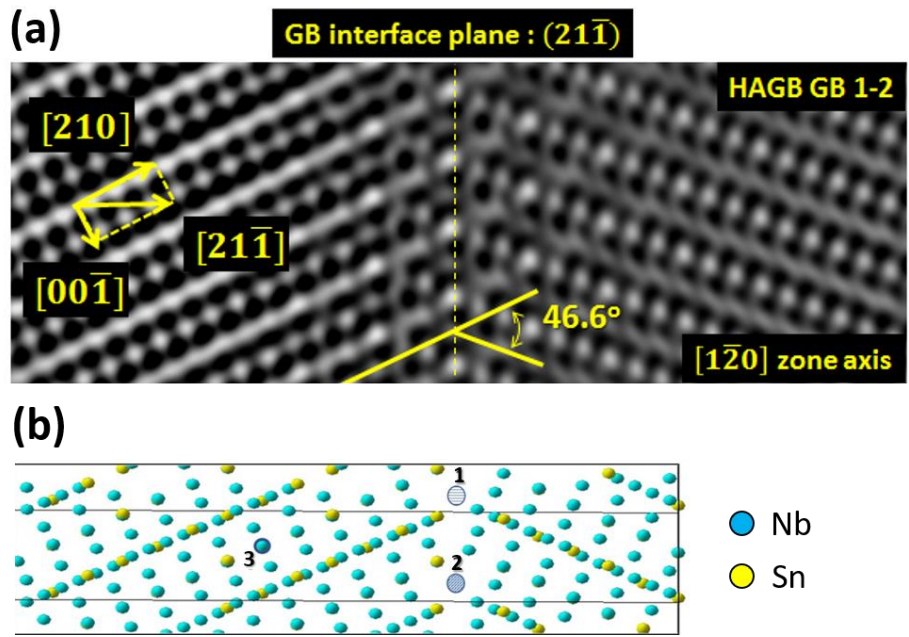

**Figure 8.** (a) Filtered atomic-resolution HAADF-STEM micrograph of a 46.6°/$[1\bar{2}0]$tilt-GB of $Nb_3Sn$ whose GB plane is $(21\bar{1})$. (b) The $[1\bar{2}0]$ tilt GB structural model determined from first-principles calculations. Site 1 or 2 is the Sn occupation site with a larger free volume space, and site 3 is an Sn anti-site in the $Nb_3Sn$ bulk. The GB structure was fully relaxed.

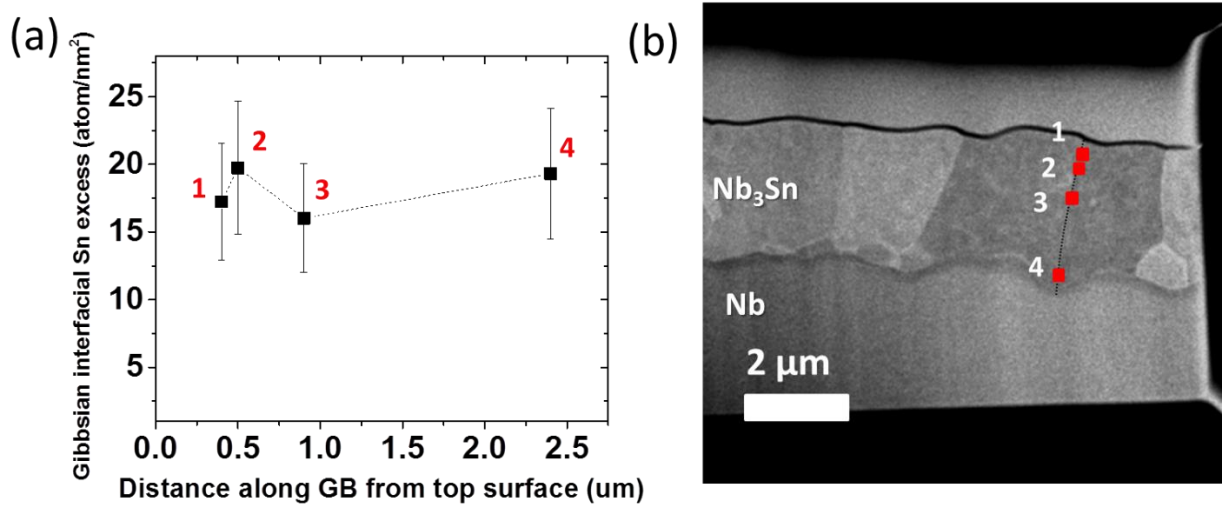

**Figure 9.** Gibbsian interfacial excesses of Sn at a Nb$_3$Sn GB along the depth direction. The values of the Gibbsian interfacial excesses of Sn are approximately a constant along the GB from the surface (point 1) to the Nb substrate (point 4), indicative of fast short-circuit diffusion of Sn along this GB, in the absence of a significant Sn concentration gradient.

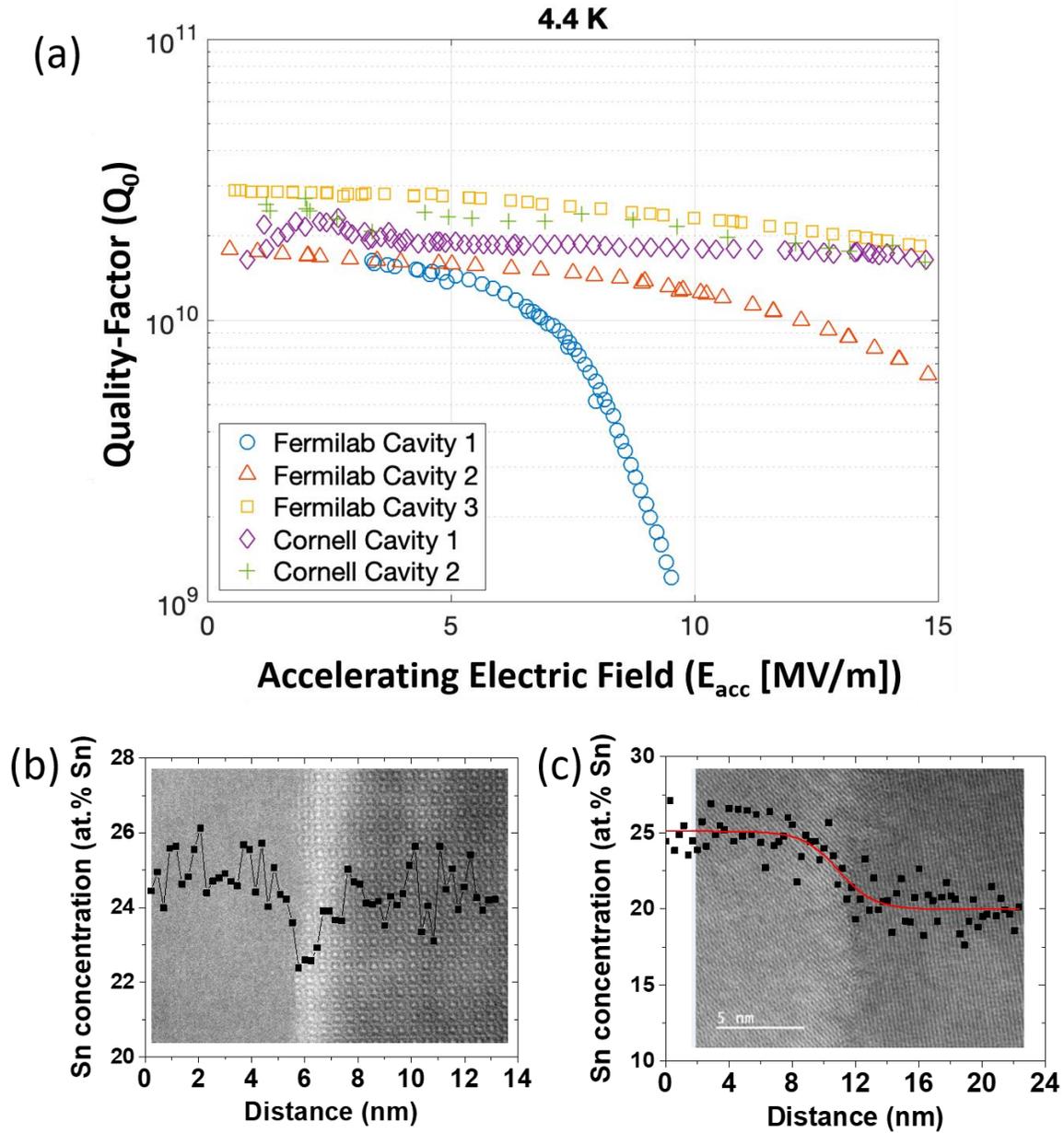

**Figure 10**. (a) Quality factor ($Q_0$) versus accelerating electric field ($E_{acc}$) curves of $Nb_3Sn$ SRF cavities measured at 4.4 K. (b) and (c): Tin concentration profiles across GBs for a high-performance $Nb_3Sn$ SRF cavities measured by STEM-EDS; (b) Cornell cavity 2 and (c) Fermilab cavity 3. We note that the high-performance cavities, Cornell cavity 2 and Fermilab cavity 3 reveal no Sn segregation at GBs.

For the 3-D reconstructions in **Figs 2** and **3**, all the isotopes of Nb and Sn are selected including the oxide states for Nb-O$^+$ and Nb-O$^{2+}$. The full width of each peak is selected for the reconstructions except Nb$^{2+}$ and Nb$^{3+}$, which have wide peak spreads. For nitrogen, N$^+$ and N$_2^+$ are selected for the 3-D reconstructions.

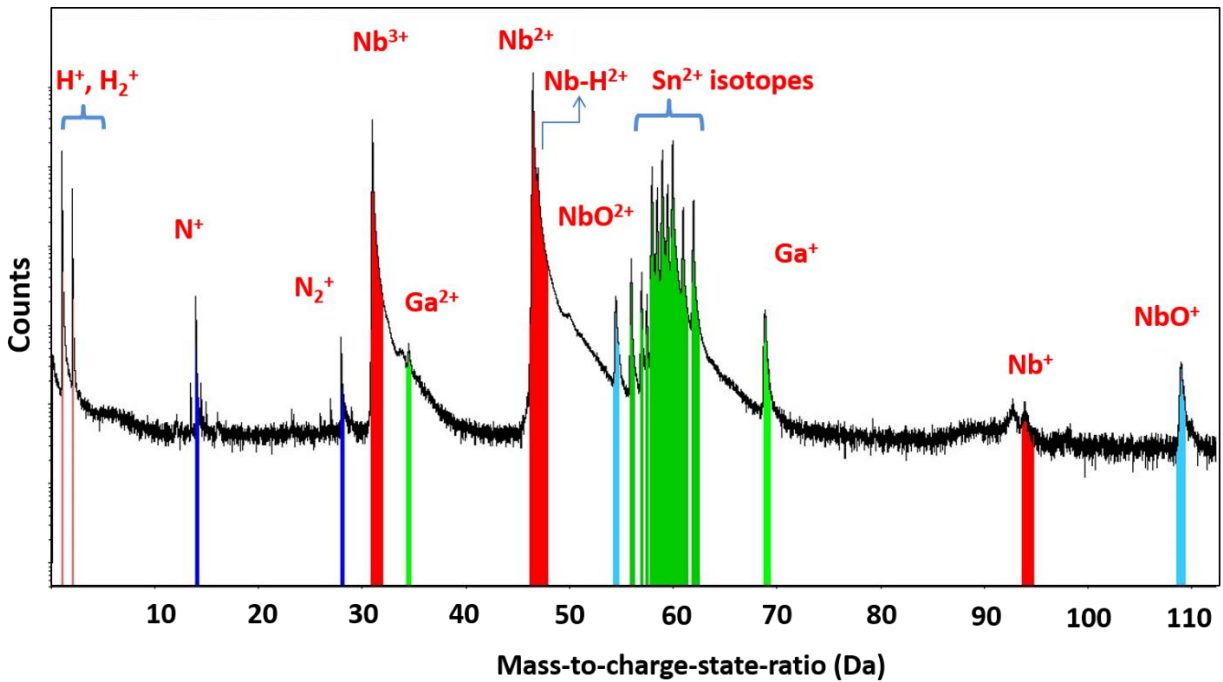

**Figure S.1** Mass spectrum for atom-probe tomographic analyses of a Nb$_3$Sn sample (A5)

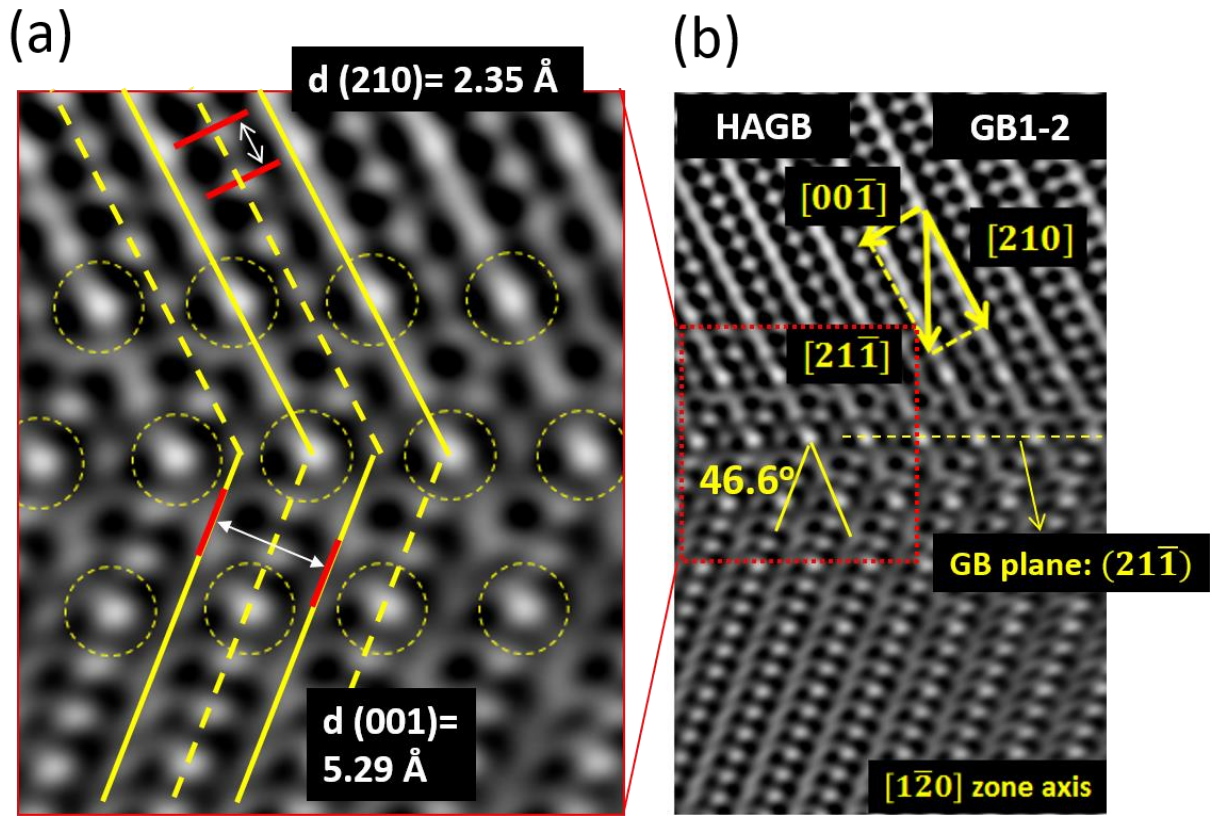

**Figure S.2** (a) Magnified filtered high-resolution high angle annular dark field (HAADF) scanning transmission electron microscopy (STEM) image of a high-angle (46.6º) [1$\bar{2}$0] tilt GB of Nb$_3$Sn. (b) The atomic planes are idexed and the GB plane is given by **n** = [21$\bar{1}$].

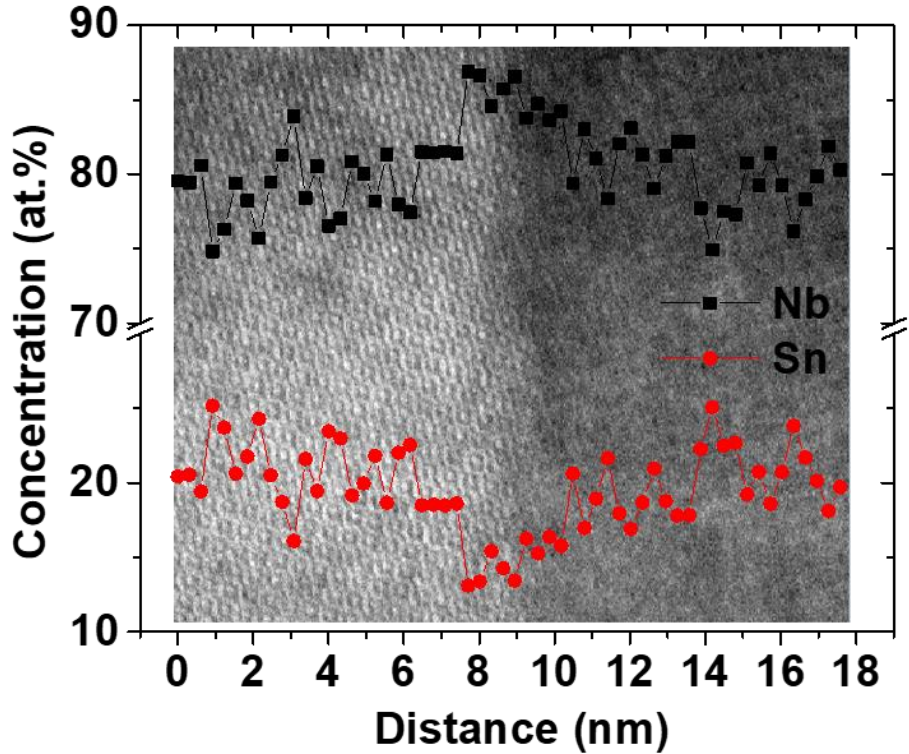

**Figure S.3** Niobium (Nb) and tin (Sn) concentration profiles across another GB of an annealed $Nb_3Sn$ sample at 1100 °C for 3 h. This figure demonstrates that when Sn segregation disappears then Nb segregation (20 ± 3 atom/$nm^2$) appears at the same GB with a full width of ~5 nm.

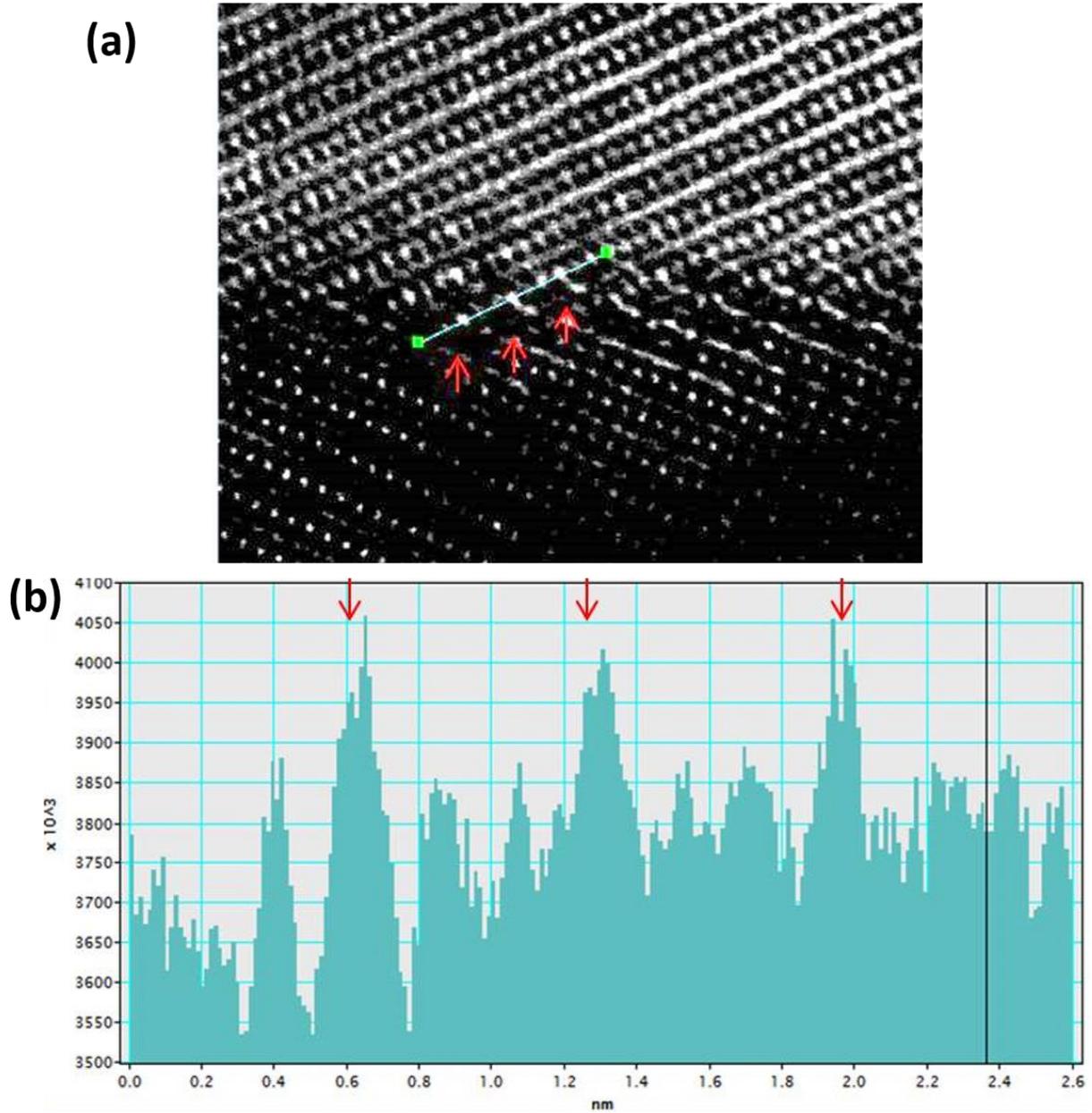

**Figure S.4** (a) high resolution HAADF- STEM image and contrast profiles across a 46.6°/[1$\bar{2}$0]-tilt-GB of $Nb_3Sn$. Bright contrast effects are observed at the coincident site lattice (CSL) sites of the GB, indicating possible chemical ordering at this $Nb_3Sn$ GB. (b) The intensity profile across CSL sites near the GB displays high intensities at the CSL.

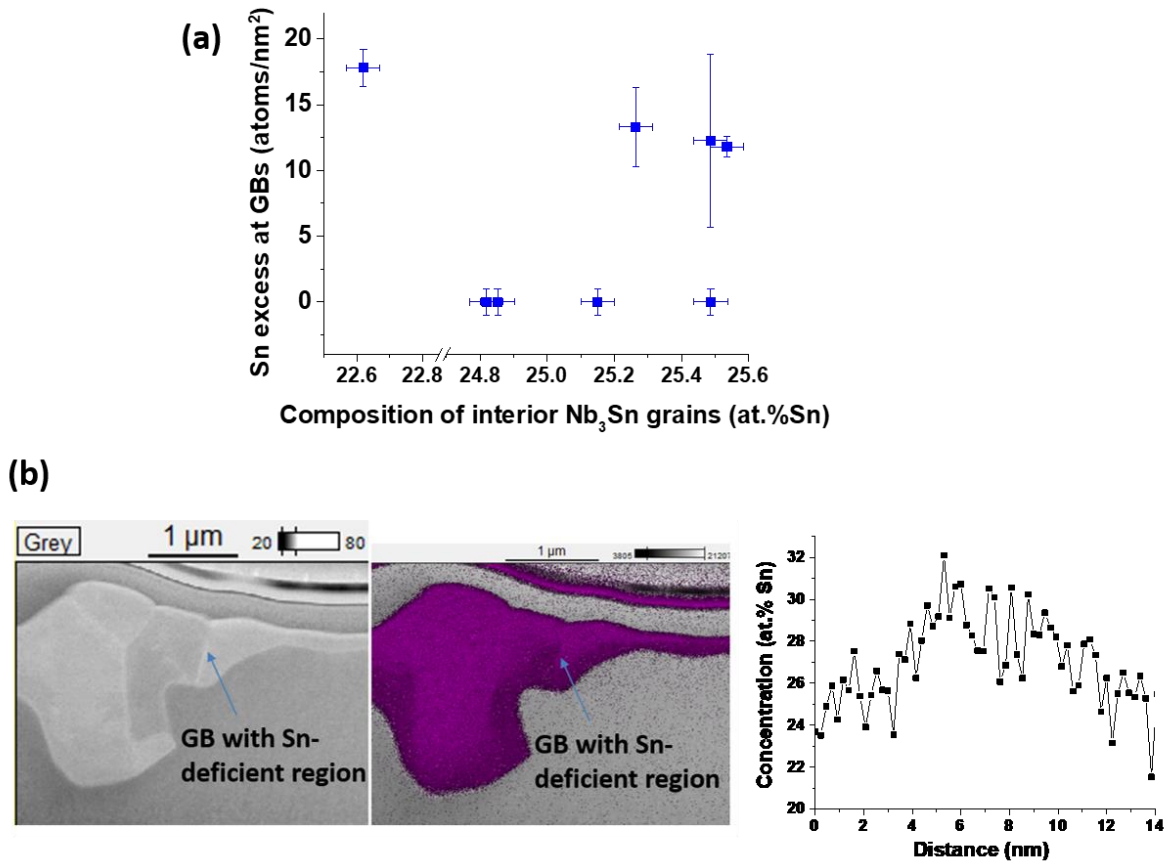

**Figure S.5** (a) The compositions of interior Nb$_3$Sn grains with different levels of interfacial excesses of Sn at GBs. Each data point represents each coating: Tin excess at GBs on the ordinate scale is an average of Sn segregation effects at 1 to 3 GBs in each coating, estimated by HR-STEM-EDS measurements. The composition of the interior Nb$_3$Sn grains on the abscissa is an average of 1 to 3 Nb$_3$Sn grains of each coating, estimated by APT analyses. The limitation of this plot is that it it is difficult to think that the composition of 2-3 Nb$_3$Sn grains represents the composition of the whole Nb$_3$Sn sample of each coating. It may provide, however, a clue indicating that there isn't a strong correlation between the composition of Nb$_3$Sn grains and Sn segregation at GBs. (b) HAADF-STEM image and EDS Sn L-map of Nb$_3$Sn grains with Sn-deficient regions. Note that a GB in partially Sn-deficient grains (<25 at.% Sn) displays Sn segregation at the GB employing an EDS measurement. A large width of the Sn segregated region (~8 nm) in the concentration profile is due to the drift of the TEM sample during data acquisition, which overestimates the Sn signal, but the existence of Sn segregation is clear.